\documentclass[prb,amsmath,amssymb,reprint,superscriptaddress]{revtex4-1}
\usepackage{amsmath}
\usepackage{amsfonts}
\usepackage{amssymb}
\usepackage{epsfig}
\usepackage{graphicx}
\usepackage{color}
\newcommand{\ave}[1]{\left\langle {#1} \right\rangle}

\begin{document}

\title{
Spurious violation of the Stokes--Einstein--Debye relation in
supercooled water
}

\author{Takeshi Kawasaki}
%\thanks{Equally contributed to this work.}
\email{kawasaki@r.phys.nagoya-u.ac.jp}
\affiliation{Department of Physics, Nagoya University, Nagoya 464-8602, Japan}

\author{Kang Kim}
%\thanks{Equally contributed to this work.}
\email{kk@cheng.es.osaka-u.ac.jp}
\affiliation{Division of Chemical Engineering, Graduate School of
Engineering Science, Osaka University, Toyonaka, Osaka 560-8531, Japan}
\affiliation{Institute for Molecular Science, Okazaki, Aichi 444-8585, Japan}

\date{\today}

\begin{abstract}
The theories of Brownian motion, the Debye rotational diffusion model,
and hydrodynamics together provide us with the Stokes--Einstein--Debye
(SED) relation between the rotational relaxation time of the $\ell$-th degree
Legendre polynomials $\tau_\ell$, and viscosity divided by temperature,
$\eta/T$. 
Experiments on supercooled liquids are frequently performed to
measure the SED relations, $\tau_{\ell}k_{\rm B}T/\eta$ and $D_{\rm t}\tau_{\ell}$, where 
$D_{\rm t}$ is the translational diffusion constant. 
However, the SED relations break down, and its
molecular origin remains elusive. 
Here, we assess the validity of the
SED relations in TIP4P/2005 supercooled water using molecular dynamics
simulations. 
Specifically, we demonstrate that the
higher-order $\tau_\ell$ values exhibit a temperature dependence similar to that
of $\eta/T$, whereas the lowest-order $\tau_\ell$ values are decoupled with $\eta/T$,
but are coupled with the translational diffusion constant. 
We reveal that the SED relations are so spurious that they
significantly depend on the degree of Legendre polynomials.
\end{abstract}

\maketitle

%\section{Introduction}

Characterization of the translational and
rotational motions of molecules in liquid states is of great
significance.~\cite{Berne:2000wf, Bagchi:2012tz, Hansen:2013uv}
For this purpose,
various transport properties, such as shear viscosity,
translational diffusion constant, and rotational relaxation time have been
measured both experimentally, and through molecular dynamics (MD) simulations.
These properties play crucial roles in
the understanding of the detailed mechanism of hydrogen-bond network
dynamics.~\cite{Conde:1983ez, Teixeira:1985dv, Agmon:1996jx,
Laage:2006jw, Laage:2008he, Moilanen:2008kb, Qvist:2011ju,
Qvist:2012gg}

The Stokes--Einstein (SE) relation
is one of the important characteristics of the translational diffusion
constant, $D_\mathrm{t}$, in many liquid state systems,
$
D_\mathrm{t} = k_\mathrm{B}T/(6\pi \eta R),
$
where $k_\mathrm{B}$, $T$, $\eta$ represent the Boltzmann constant, the
temperature, and the shear viscosity, respectively.
This SE relation is derived originally from the theories of hydrodynamics
and Brownian motion, where a rigid spherical particle with a radius
$R$ is assumed to be perfectly suspended in a Stokes flow of a
constant shear viscosity $\eta$ under the stick boundary condition.~\cite{Landau:1987hg}
Thus, $R$ is conventionally regarded as the effective hydrodynamic
radius of the molecule when applying the SE relation to molecular liquids.~\cite{Schmidt:2003fd}

Analogous to translational motion, the rotational Brownian motion
leads to another SE relation between the rotational diffusion constant,
$D_\mathrm{r}$, and $\eta$ as
$
D_\mathrm{r} = k_\mathrm{B}T/(8\pi \eta R^3).
$
Based on the Debye model, $D_\mathrm{r}$ can also be determined by solving the rotational
diffusion equation for the reorientation of the molecular dipole as
$
D_\mathrm{r}  = 1/[\tau_\ell \ell(\ell+1)],
$
where $\tau_\ell$ is the rotational relaxation time of the $\ell$-th order 
Legendre polynomials.~\cite{Debye:1929tl}
Note that $\tau_1$ and $\tau_2$ are the most-commonly investigated; they are
characterized by dielectric relaxation and NMR 
spectroscopies, respectively.
A deviation from $\tau_1/\tau_2=3$ has been reported in supercooled
molecular liquids, which is regarded as a
sign of the breakdown of the Debye model.~\cite{Kivelson:1988fu, Kivelson:1989ef,
Diezemann:1998bj}
Those two equations result in the
Stokes--Einstein--Debye (SED) relation, 
\begin{align}
\frac{\tau_\ell k_\mathrm{B}T} {\eta} = \frac{8\pi R^3}{\ell(\ell+1)}.
\label{eq:tau_ell_eta}
\end{align}
The SED relation can also be expressed as
\begin{align}
D_\mathrm{t} \tau_\ell = \frac{4R^2}{3\ell (\ell+1)},
\label{eq:Dt_tau_ell}
\end{align}
by combining further with the SE relation between 
$D_\mathrm{t}$ and $\eta/T$.
This SED relation is proportional to the quotient
$D_\mathrm{t}/D_\mathrm{r}$, which accounts for the coupling
between the translational and rotational diffusion dynamics at any temperature.

The violation of the SE relation between $D_\mathrm{t}$ and $\eta$ has been intensively observed in
various glass-forming liquids, such as
\textit{o}-terphenyl.~\cite{Fujara:1992ib, Chang:1994eb, Cicerone:1996cb,
Andreozzi:1997ie, Ediger:2000ed, Mapes:2006fp}
%metallic glasses~\cite{Meyer:2003jl}, and colloidal
%glasses~\cite{Bonn:2003iv, Edmond2012Decoupling, Vivek:2017hs}.
In particular, the quantity $D_\mathrm{t}\eta/T$ increases towards
the glass transition temperature, but
exhibits a constant value at high temperatures.
These experiments indicate that the translational diffusion occurs in a more
enhanced manner than estimations using shear viscosity.
Many theoretical efforts have therefore been devoted to explaining the violation of the
SE relation in glass-forming liquids.~\cite{Hodgdon:1993ew,
Stillinger:1994ko, Stillinger:1995fu, Tarjus:1995gx,
Douglas:1998id, Xia:2001kb,
Jung:2004jz, Biroli:2007kg, Ngai:2009ge}
MD simulations have also been variously performed to address their
molecular mechanisms.~\cite{Thirumalai:1993cv,
Yamamoto:1998gb, Kumar:2006kr, Koddermann:2008jg, Harris:2009bv, Shi:2013ji, Sengupta:2013dg}
It is also commonly argued that the violation of SE relation is a sign of spatially
heterogeneous dynamics and of the non-Gaussian property of the particle
displacement distribution.~\cite{Ediger:2000ed, Berthier:2011hs}

In contrast, the validity of the SED relation is still highly controversial,
because there are three possible candidates, $\tau_\ell T/\eta$, $D_\mathrm{t}\tau_\ell$, and
$D_\mathrm{t}/D_\mathrm{r}$, that need to be quantified.
%Those three quantities are consistent with each other if the hydrodynamic radius $R$ remains
%unchanged at any temperature both for translational and rotational Brownian motions.
%However, the direct measurement of $R$ is impracticable for molecular
%liquids both in experiments and MD simulations.
%Furthermore, the concept of the rotational diffusion constant
%$D_\mathrm{r}$ itself is artificial rather than that of $D_\mathrm{t}$
%The validity of the Debye model for
%molecular reorientation, \textit{i.e.}, the $\ell$ dependence of
%$\tau_\ell$, hinders us from making a precise assessment for
%the SED relation, whatever one of three quantities are utilized.}
Recently, the $D_\mathrm{r}$ of supercooled
molecular liquids
has been calculated using MD simulations following the Einstein
relation for rotational Brownian motions.~\cite{Kammerer:1997ku, Lombardo:2006jq, Chong:2009ci}
Experimental analogs have also been reported using optical spectroscopy
in colloidal glasses.~\cite{Kim:2011he, Edmond:2012es}
In particular, it has been shown that the temperature dependences of
$D_\mathrm{t}\tau_2$ and $D_\mathrm{t}/D_\mathrm{r}$ are completely
different in \textit{o}-terphenyl liquids~\cite{Lombardo:2006jq} and diatomic
molecular liquids~\cite{Chong:2009ci};
$D_\mathrm{t}/D_\mathrm{r}$ significantly decreases with decreasing 
temperature, indicating the translational-rotational decoupling.
In contrast, $D_\mathrm{t}\tau_2$ exhibits the opposite temperature
dependence, \textit{i.e.}, increases in $\tau_2$ exceed the time scale of
the translational diffusion constant, $1/D_\mathrm{t}$, as 
the temperature decreases.
This discrepancy is thus attributed to the inconsistency between the two expressions,
Eqs.~(\ref{eq:tau_ell_eta}) and (\ref{eq:Dt_tau_ell}).
However, the direct measurement of $R$ is impractical for molecular
liquids both in experiments and MD simulations.
More practically, the breakdown of the Debye model, \textit{i.e.}, the $\ell$ dependence of
$\tau_\ell$, prevents us from making a precise assessment of
the SED relation, whichever one of three quantities is utilized.

For liquid water, 
it has been widely accepted that the validity of 
the Debye model for molecular reorientation is limited even in normal
states, although that is frequently used when analyzing experimental data.
Instead, various large-amplitude rotational jump models have been developed
to give an accurate prediction of the rotational relaxation time
$\tau_2$.~\cite{Laage:2006jw, Laage:2008he, Qvist:2012gg}
Particularly for supercooled water, the appropriate description for the violation
of the SED relation becomes more complicated.
Recent MD simulations have demonstrated
that the translational and rotational dynamics become
spatially heterogeneous upon cooling.~\cite{Giovambattista:2004ft,
Mazza:2006bk}
Furthermore, the violations of the SE and SED relations have been intensely 
characterized through both experiments and simulations.~\cite{Chen:2006kk,
Becker:2006ju, Kumar:2006hx, Kumar:2007hl, Mazza:2007kr,
Xu:2009hq, Banerjee:2009db, Mallamace:2010uj, Jana:2011fj, Qvist:2012gg,
Rozmanov:2012ja, Bove:2013em, Dehaoui:2015ii,
Guillaud:2017bk, Guillaud:2017ey,
Galamba:2017eq, Kawasaki:2017gw, Shi:2018gu, MonterodeHijes:2018ec,
Saito:2018dn, Saito:2019ht}
In particular, the violation of the SED relation and the 
translational-rotational decoupling in supercooled water have been reported
by calculating $D_\mathrm{t}$ and $D_\mathrm{r}$, while
$\eta$ has not been calculated~\cite{Becker:2006ju, Mazza:2007kr,
Rozmanov:2012ja}, 
despite the fact that $\eta$ plays an essential role
in the precise assessment of the SE relation.~\cite{Guillaud:2017bk,
Guillaud:2017ey, Galamba:2017eq, Kawasaki:2017gw}
The SED relation has been investigated by calculating the $\eta$ of SPC/E
supercooled water, during which
$D_\mathrm{r}$ was not calculated.~\cite{Galamba:2017eq}
Under these conditions, the SED relation, particularly for the $\ell$
dependence of $\tau_\ell T/\eta$, has not yet been thoroughly
investigated, while only one experimental data analysis has been conducted for
$\tau_2T/\eta$.~\cite{Dehaoui:2015ii}

The purpose of this study is to shed light on the controversy regarding
the violation of the SED relation, specifically through the numerical calculations
of three quantities, $\tau_\ell T/\eta$, $D_\mathrm{t}\tau_\ell$, and
$D_\mathrm{t}/D_\mathrm{r}$.
In particular, we aim to demonstrate that the $\ell$ dependence of $\tau_\ell
T/\eta$ is an important factor in exploring the inherent
translational-rotational dynamics in supercooled water.

%01
\begin{figure*}[t]
\centering
\includegraphics[width=0.85\textwidth]{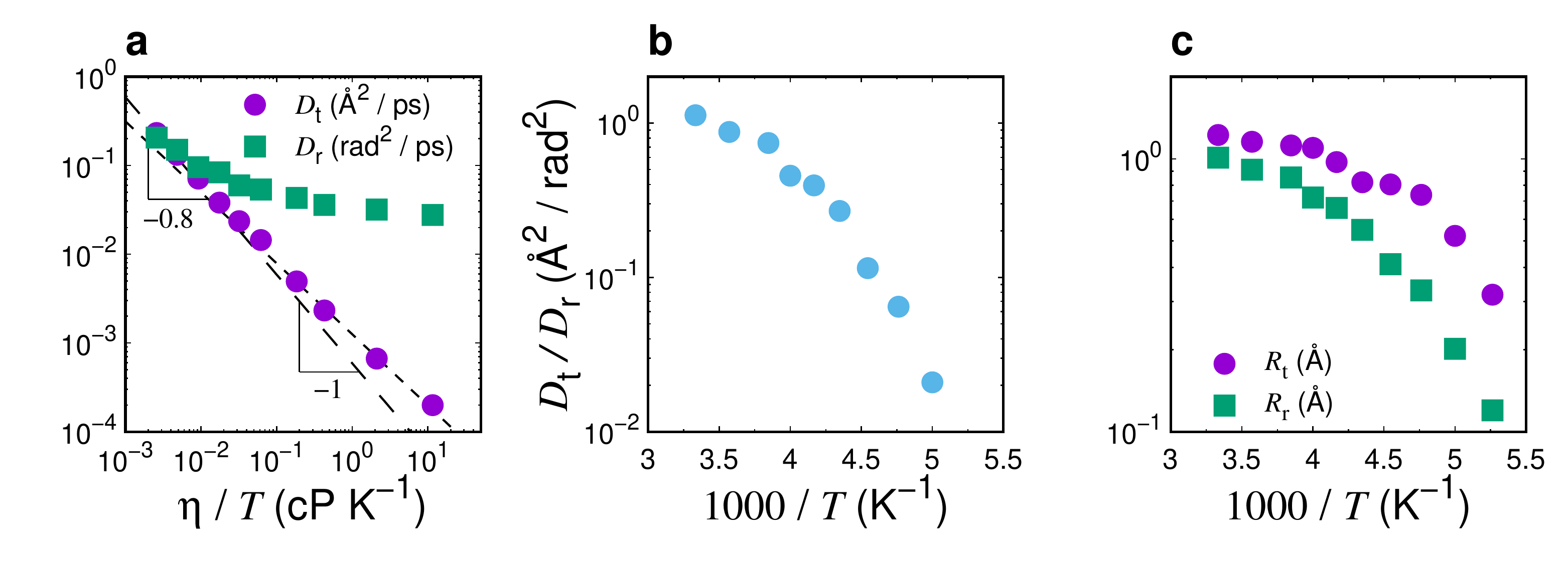}
\caption{\label{fig1} 
(a) 
Assessments of the translational and rotational SE relations, $D_\mathrm{t}\eta/T$ and
 $D_\mathrm{r}\eta/T$, made by plotting the 
relationships between the translational diffusion constant $D_\mathrm{t}$
 or the rotational diffusion constant $D_\mathrm{r}$ and the shear viscosity
 divided by the temperature $\eta/T$. 
Long-dashed line represents the
 linear relation $D \propto \eta/T$, which represents the SE
 relation. 
Short-dashed line denotes the fractional SE relation, $D_\mathrm{t}\propto (\eta/T)^{-0.8}$.
Neither the translational nor the rotational SE relations are 
 satisfied in supercooled region ($T < 250$ K). 
(b) Temperature dependence of the ratio of 
 rotational and translational diffusion constants, $D_\mathrm{r}/D_\mathrm{t}$.
As $T$ decreases, this ratio increases, indicating the
 translational-rotational diffusion decoupling.
(c) Temperature dependence of translational and rotational hydrodynamic radii,
 $R_\mathrm{t}$ and $R_\mathrm{}$.
Both $R_\mathrm{t}$ and $R_\mathrm{r}$ decrease
significantly upon cooling, accompanied with violation of SE
 relations. 
In particular, upon cooling, $R_\mathrm{r}$ decreases at a higher rate than
 that of $R_\mathrm{t}$ in response to decreasing temperature.
}
\end{figure*}

%02
\begin{figure*}[t]
\centering
\includegraphics[width=0.85\textwidth]{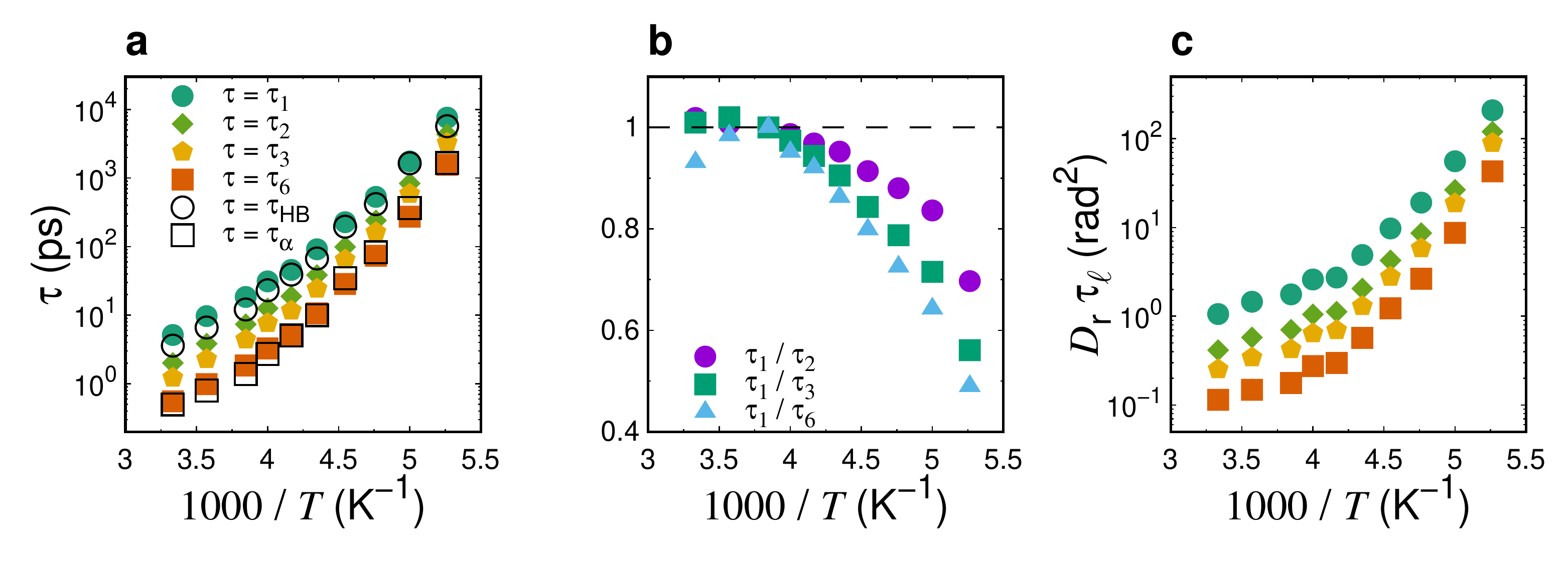}
\caption{\label{fig2} 
(a) Temperature dependence of $\ell$-th
 order rotational relaxation times $\tau_{\ell}$ for $\ell=1$, $2$ ,$3$, and
 $6$. 
The temperature dependences of hydrogen-bond lifetime,
 $\tau_\mathrm{HB}$, and $\alpha$-relaxation time, $\tau_\alpha$, are
 included for comparison.
(b) Temperature dependence of ratio $\tau_1/\tau_2$, $\tau_1/\tau_3$,
 and $\tau_1/\tau_6$.
Each quantity is scaled by the value at $T = 260$ K. 
(c) 
Assessments of the Debye model, made by plotting the
 temperature dependence of $D_\mathrm{r}\tau_{\ell}$ for
 $\ell=1$, $2$, $3$, and $6$. 
}
\end{figure*}

\section*{Results}
%%%%Fig1%%%%%%%%%%%%%%%%
Here we examine the translational and rotational SE relations,
$D_\mathrm{t}\propto \eta/T$ and $D_\mathrm{r}\propto \eta/T$, respectively.
We determined
$D_\mathrm{t}$ and $D_\mathrm{r}$ from the long-time behaviors of 
the translational and rotational mean-square displacements, respectively (see
Methods).
We calculated $\eta$ using the Green--Kubo formula
for the shear stress correlation function, as detailed in a previous study.~\cite{Kawasaki:2017gw}
Figure~\ref{fig1}(a) shows both $D_\mathrm{t}$ and $D_\mathrm{r}$ as a function
of $\eta/T$. 
Comparing these with the dashed line representing the linear relationship,
we find that both the translational and the rotational SE relations
are invalid in supercooled regimes, particularly at $T<250$ K.
Note that the rotational SE relation is violated to a greater extent than 
the translational SE relation. 
Figure~\ref{fig1}(b) shows the ratio of the
translational and rotational diffusion constants,
$D_\mathrm{t}/D_\mathrm{t}$, as a
function of the scaled inverse of temperature. 
The substantial decoupling displayed between the two diffusion constants
indicates that 
the translational and rotational dynamics are decoupling, which is comparable with the
previously reported results on ST2~\cite{Becker:2006ju},
SPC/E~\cite{Mazza:2007kr}, and TIP4P/2005~\cite{Rozmanov:2012ja} models.
Furthermore, similar results are also demonstrated
in \textit{o}-terphenyl liquids~\cite{Lombardo:2006jq} and diatomic
molecular liquids~\cite{Kammerer:1997ku, Chong:2009ci} using MD simulations.

The observed decoupling of translational-rotational diffusion is directly related
to the inconsistency regarding the effective hydrodynamic radius
observed when using the SE relations.
We quantified the hydrodynamic radius for the translational
degree of freedom, $R_\mathrm{t}= k_\mathrm{B}T/(6\pi\eta
D_\mathrm{t})$, and the rotational counterpart,
$R_\mathrm{r}=[k_\mathrm{B}T/(8\pi\eta D_\mathrm{r})]^{1/3}$.
Figure~\ref{fig1}(c) shows the temperature dependences of 
$R_\mathrm{t}$ and $R_\mathrm{r}$.
At $T=300$ K, $R_\mathrm{t}$ and $R_\mathrm{r}$ 
are approximated by $1.2$~\AA~and $1.0$~\AA, respectively.
These values are slightly smaller than the van der Waals radius of
the TIP4P/2005 model.
As seen in Fig.~\ref{fig1}(c), these two radii sharply decrease upon supercooling, 
accompanied by violation of the translational and rotational SE relations.
Moreover, the difference between $R_\mathrm{t}$
and $R_\mathrm{r}$ increases with decreasing the temperature, implying
that the translational and rotational diffusions are decoupling.
The relevance of the decoupling $D_\mathrm{t}/D_\mathrm{r}$ will be discussed below.

%%%%Fig2%%%%%%%%%%%%%%%%
Next, we investigate $\tau_\ell$ as determined from the $\ell$-th
order Legendre polynomials, and explore its relationship with $D_\mathrm{r}$.
Figure~\ref{fig2} (a) shows
$\tau_{\ell}$ (for $\ell =1$, $2$, $3$, and $6$) as a function of the scaled
inverse of the temperature.
$\tau_{\ell}$ increases for all
$\ell$ values as the temperature decreases.
Interestingly, we observe that $\tau_{\ell}$ values with higher-order degrees
exhibit stronger temperature dependence than those of the lowest order.
In other words, the ratios $\tau_1/\tau_2$, $\tau_1/\tau_3$, and
$\tau_1/\tau_6$ notably decrease as the temperature decreases (see Fig.~\ref{fig2}(b)).
A similar result was found using MD simulations of the SPC/E supercooled water.~\cite{Galamba:2017eq}
%This result is direct evidence of the violation of the Debye model for
%rotational diffusion, $\tau_\ell=1/[D_\mathrm{r}\ell(\ell+1)]$.
As evident in Fig.~\ref{fig2}(b),
$D_\mathrm{r}\tau_\ell$ exhibits strong temperature
dependence, indicating the breakdown of the Debye model.
The observed deviation increases for higher-order $\ell$ values with decreasing temperatures.
%%%%%%Fig3%%%%%%%%%%%%%%%%%%%%%%%%%%
The breakdown of the Debye model
and the inconsistency between $R_\mathrm{t}$ and $R_\mathrm{r}$ in
supercooled states suggest that 
the SED relations, $\tau_\ell T/\eta$, $D_\mathrm{t}\tau_\ell$, and
$D_\mathrm{t}/D_\mathrm{r}$, are likely spurious quantities.
More precisely, these quantities cannot represent real 
translational and rotational dynamics in supercooled water, regardless
of whether they 
exhibit anomalous deviations from values at high temperatures.
We below demonstrate ambiguities of the SED relations, of which results
markedly depend on the order $\ell$.

\begin{figure}[t]
\centering
\includegraphics[width=0.5\textwidth]{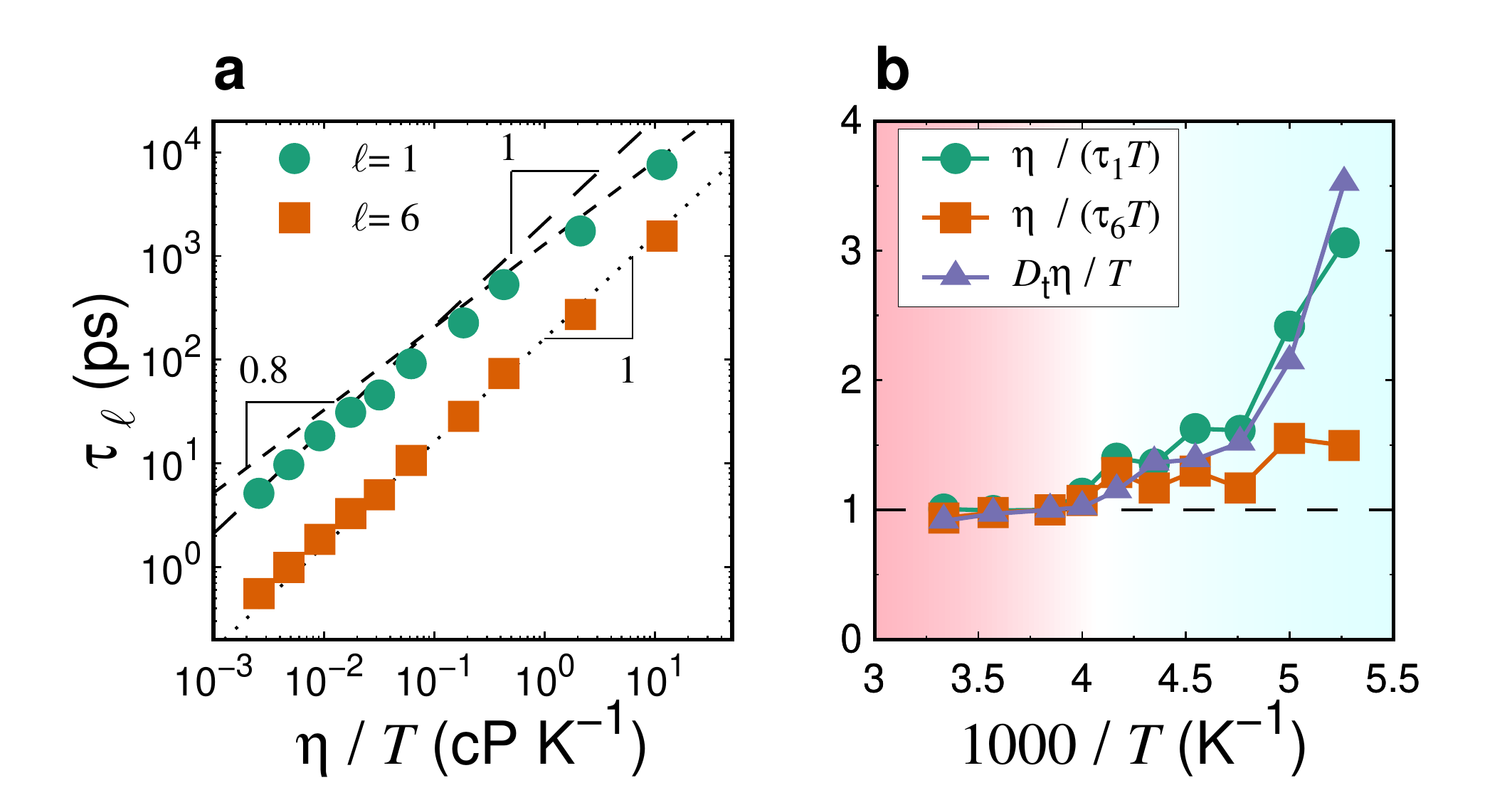}
\caption{\label{fig3} 
(a) Assessments of the SED relation, $\tau_{\ell}T/\eta$, made by plotting
the relationships between rotational relaxation time $\tau_{\ell}$ for
 $\ell=1$ and $\ell=6$, and the shear viscosity divided by
 the temperature, $\eta/T$.
Both the dotted line and the long-dashed
 line represent the SED relation, $\tau_{\ell}\propto \eta/T$. 
The short-dashed line represents $\tau_{\ell}\propto (\eta/T)^{0.8}$. 
(b) Assessments of the SED relation, $\tau_{\ell}T/\eta$, made by plotting
 the temperature dependence of $\eta/(\tau_\ell T)$ for
 $\ell=1$ and $\ell=6$.
The violation of the SE relation, $D\mathrm{t}\eta/T$, is also plotted
 for comparison.
Each quantity is scaled by the value at $T = 260$ K. 
The SED ratio $\eta/(\tau_1 T)$ exhibits a 
temperature dependence similar to the violation of the SE relation, $D_\mathrm{t} \eta/T$, whereas
the plot of $\eta/(\tau_6 T)$ resembles the SED relation, Eq.~(\ref{eq:tau_ell_eta}).
The background color (white region) indicates the onset temperature of
 the SE violation.
}
\end{figure}

Here, we address the SED relation $\tau_\ell T/\eta$, which is the 
counterpart to recent experimental data.~\cite{Dehaoui:2015ii}
Figure~\ref{fig3}(a) shows the relationship between 
$\eta/T$ and $\tau_{\ell}$ for $\ell=1$
and $\ell=6$.
Note that the results of $\ell=2$ and $\ell=3$ are omitted from the plot
to improve its clarity.
As observed in Fig.~\ref{fig3}(a), $\tau_{1}$ deviates from the value
predicted by the SED relation, particularly at lower temperatures
($T < 250$ K), instead exhibiting the fractional form $\tau_1 \propto (\eta/T)^{-0.8}$.
In contrast, $\tau_\ell$ with at higher-order $\ell=6$ follows the SED
relation, $\tau_6\propto \eta/T$.
Figure~\ref{fig3}(b) shows the temperature dependence of
$\eta/(\tau_{\ell }T)$ (for $\ell=1$ and $\ell=6$), in
comparison with that of the translational SE relation, $D_\mathrm{t}\eta/T$.
We observe that the temperature dependence of 
$\eta/(\tau_{1}T)$ is analogous to that of $D_\mathrm{t}\eta/T$,
suggesting the violation of the SE relation, whereas $\eta/(\tau_{6}T)$
exhibits a weaker temperature dependence.
We previously demonstrated the relationship
$ \tau_{\alpha}\propto\eta/T$ in TIP4P/2005 supercooled
water.~\cite{Kawasaki:2017gw}
Here, $\tau_\alpha$ denotes the $\alpha$-relaxation time that was
determined from the incoherent intermediate scattering function $F_s(k, t)$.
The wave-number, $k$, was chosen as $k=3.0$ \AA$^{-1}$, which corresponds to
the main peak of the static structure factor of oxygen, $S_\mathrm{OO}(k)$.
This implies that $D_\mathrm{t}\tau_\alpha$ is a good proxy for the
translational SE relation $D_\mathrm{t}\eta/T$.
Similar results have also been reported for other supercooled
liquid systems.~\cite{Yamamoto:1998gb, Shi:2013ji, Sengupta:2013dg, Kawasaki:2014ky}
Accordingly, 
the temperature dependence of $\tau_{\ell}$ with
higher-order $\ell$ resembles the coupling with that of $\tau_{\alpha}$
(see Fig.~\ref{fig2}(a)).
On the other hand, the deviation of $\eta/(\tau_{1}T)$ from this value at
high temperatures superimposes the violation of the translational SE relation,
$D_\mathrm{t}\eta/T$ or $D_\mathrm{t} \tau_\alpha$.

%03
%04
\begin{figure}[t]
\centering
\includegraphics[width=0.5\textwidth]{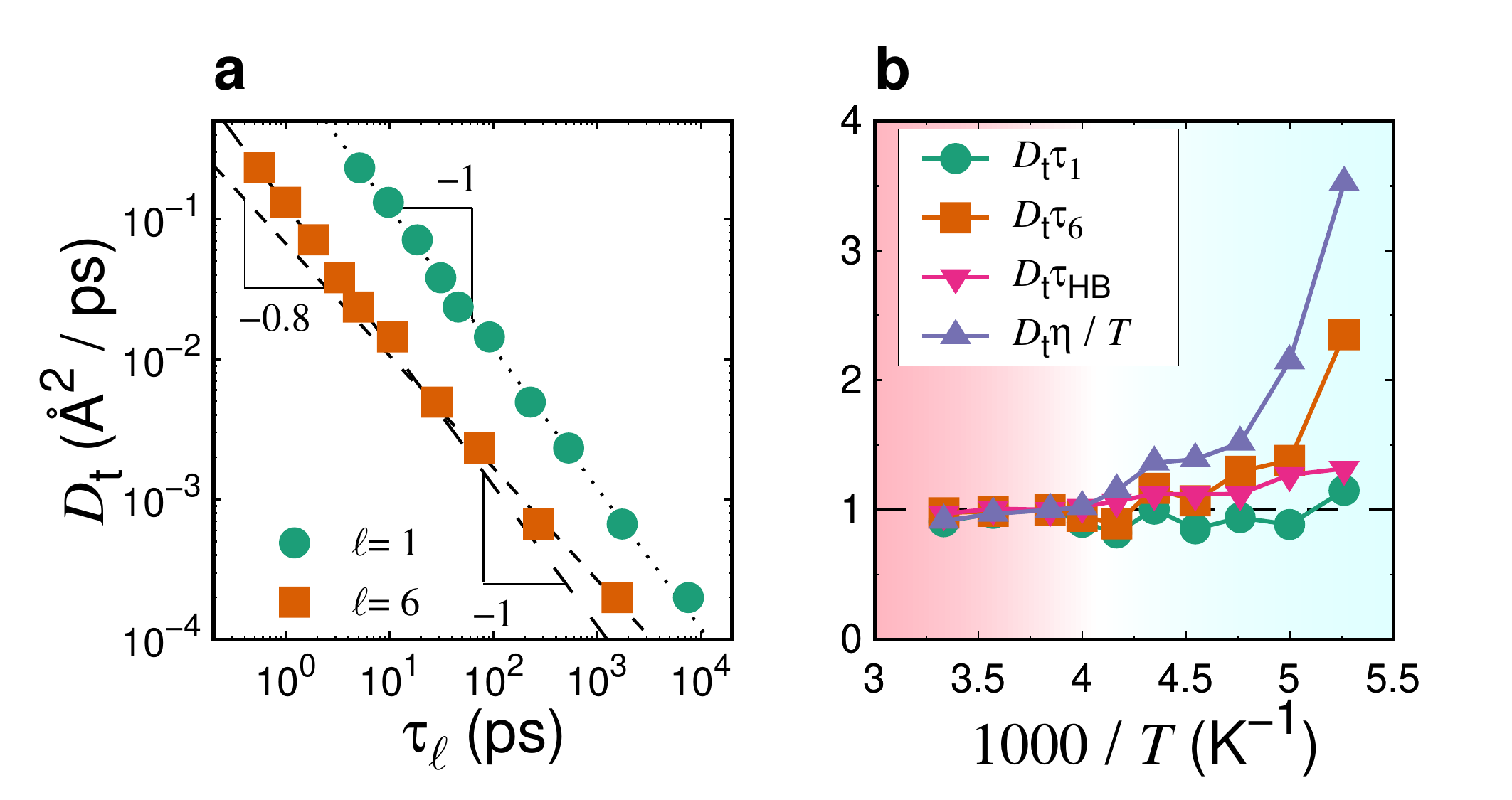}
\caption{\label{fig4} 
(a)
Assessments of the SED relation, $D_\mathrm{t}\tau_\ell$, made by plotting
the relationship between the translational diffusion constant, $D_\mathrm{t}$, and
the rotational relaxation time, $\tau_\ell$, for $\ell=1$ and $\ell=6$.
Both the dotted line and the long-dashed
 line represent the SED relation, $D_\mathrm{t}\propto \tau_{\ell}^{-1}$. 
The short-dashed line represents $D_\mathrm{t}\propto \tau_{\ell}^{-0.8}$. 
(b) 
Assessments of the SED relation, $D_\mathrm{t}\tau_\ell$, made by plotting
the temperature dependence of the SED ratios $D_\mathrm{t}\tau_\ell$.
The violation of SE relation, $D\mathrm{t}\eta/T$, is also plotted for comparison.
$D_\mathrm{t}\tau_\mathrm{HB}$, with hydrogen-bond
 lifetime $\tau_\mathrm{HB}$, is also shown.
Each quantity is scaled by the value at $T = 260$ K. 
Note that $D_\mathrm{t}\tau_\mathrm{HB}$ shows the preservation of the SE
 relation.~\cite{Kawasaki:2017gw}
The SED ratio $\eta/(\tau_1 T)$ exhibits a 
temperature dependence similar to the preservation of the SE relation, $D_\mathrm{t}\tau_\mathrm{HB}$, whereas
 $D_\mathrm{t}\tau_6$ bears a certain resemblance to the violation of
 the SE relation, $D_\mathrm{t}\eta/T$.
The background color (white region) indicates of the onset temperature
 of the SE violation.}
\end{figure}
%05
\begin{figure*}[t]
\centering
\includegraphics[width=0.85\textwidth]{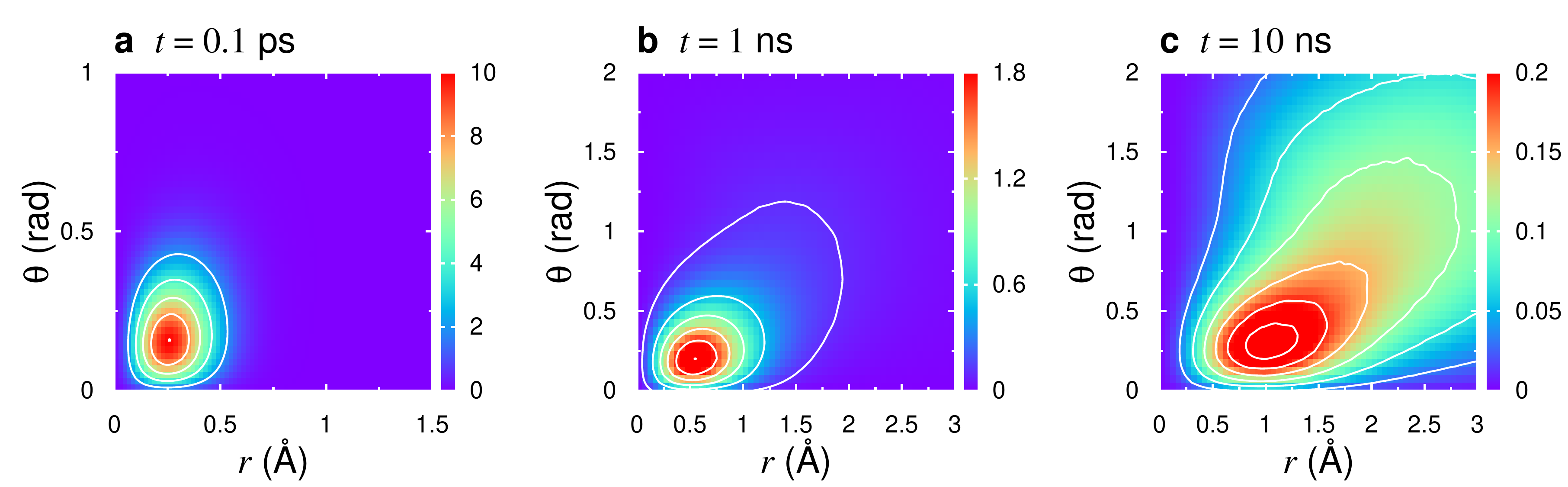}
\caption{\label{fig5} 
Joint probability distribution functions (generalized van Hove correlation function)
 $4\pi r^2 G_s(r, \theta; t)$ for $t=0.1$ ns, 1 ns, and 10 ns at
 $T=190$ K. 
The value of the color bar is normalized by \AA$^{2}$.
Remarkable positive correlations between translational
 displacement $|\Delta \vec{r}_j(t)|$ and rotational angle
 $\Delta \theta_j(t)$ are observed, particularly for large $r$ and
 $\theta$ values.
This indicates
 that a large translational motion correlates with a large rotational 
 motion of molecule.
}
\end{figure*}

%%%%%%Fig4%%%%%%%%%%%%%%%%%%%%%%%%%%
We next examine the second SED relation, $D_\mathrm{t}\tau_\ell$.
Figures~\ref{fig4}(a) and (b) display the relationship between $D_\mathrm{t}$
and $\tau_\ell$ and the temperature dependence of $D_\mathrm{t}\tau_\ell$, respectively.
As $\tau_6$ can serve as a proxy of $\eta/T$ as observed in Fig.~\ref{fig3}, 
$D_\mathrm{t}\tau_6$ exhibits a comparable temperature dependence with
the SE ratio $D_\mathrm{t}\eta/T$ (see Fig.~\ref{fig4}(b)).
In contrast, $\tau_1$ exhibits a temperature dependence similar to
that of $D_\mathrm{t}$.
%These results are all consistent with the results in
%Fig.~\ref{fig3} ($\tau_1 \propto D_\mathrm{t}^{-1}$ and $\tau_6 \propto
%\eta/T$).
Furthermore, we show an alternative quantity,
$D_\mathrm{t}\tau_\mathrm{HB}$, with the hydrogen-bond lifetime,
$\tau_\mathrm{HB}$.
Here, $\tau_\mathrm{HB}$ represents the time scale characterizing
the irreversible hydrogen-bond breakage process, which is determined
from the hydrogen-bond correlation function.~\cite{Rapaport:1983ib,
Saito:1995gz, Luzar:1996gw,
Luzar:1996gx, Luzar:2000gv}
As demonstrated in a previous study and displayed in
Fig.~\ref{fig4}(b), 
the SE relation is preserved as $D_\mathrm{t}\sim {\tau_\mathrm{HB}}^{-1}$ if
the time scale $\tau_\alpha$ is replaced with
$\tau_\mathrm{HB}$.~\cite{Kawasaki:2017gw}
Similar observations have also been reported in binary soft-sphere supercooled
liquids~\cite{Kawasaki:2013bg} and silica-like network-forming supercooled liquids.~\cite{Kawasaki:2014ky}
This SE preservation is understood by a possible ``jump model.'' 
As outlined in the Ref.~\citenum{Kawasaki:2017gw}, the frequency of
the jump motion can be represented as $f\sim 1/\tau_\mathrm{HB}$ at investigated temeperatues.
Correspondingly, the translational diffusion constant is modeled as
$D_\mathrm{t}\sim \ell_\mathrm{jump}^2f\sim
\ell_\mathrm{jump}^2/\tau_\mathrm{HB}$, where $\ell_\mathrm{jump}$
denotes a typical jump length ($\sim$~\AA).
Therefore, $D_{\rm t}\tau_\mathrm{HB}$ becomes constant.
This SE preservation indicates that 
irreversible hydrogen-bond breakages destroy the local tetrahedral
structures, and lead to the translational and rotational molecular jumps
with high mobility.
It also implies the
the coupling between $\tau_1$ and $\tau_\mathrm{HB}$, which is
demonstrated in Fig.~\ref{fig4}.

%%%%%%Fig5%%%%%%%%%%%%%%%%%%%%%%%%%%
To elucidate the molecular mechanism of the demonstrated relationship
between $D_\mathrm{t}$ and $\tau_\ell$, we analyze the generalized van
Hove self correlation function, \textit{i.e.}, the joint probability
distribution function for the translational
displacement and the rotational angle of the molecule, $G_s(\vec{r}, \theta; t) =
(1/N)\sum_{j=1}^N  \ave{\delta(\vec{r}-\Delta
\vec{r}_j(t))\delta(\theta-\Delta \theta_j(t))}$.~\cite{Faraone:2003fm}
Here, $\Delta \vec{r}_j(t)$ and $\Delta \theta_j(t)$ are the translational
displacement vector of oxygen and the rotational angle of the dipole moment
of a molecule $j$ during a time $t$, respectively.
Figure~\ref{fig5} shows the contour maps of $4\pi r^2 G_s(r, \theta; t)$
with $r=|\vec{r}|$ for $t=0.1$ ps, 1 ns, and 10 ns at $T=190$ K.
For the shorter time interval, $t=0.1$ ps, the distribution is stretched towards the
rotational angle direction, $\theta$, which is caused by the libration
motion of the molecule.
This observation corresponds to 
the oscillations of $C_\ell(t)$ (see Supplementary Fig.~S2).
At longer time scales, however,
$G_s(r, \theta; t)$ shows the coupling between the translational displacement and
the rotational angle, which is consistent with the previously reported
results of Ref.~\citenum{Faraone:2003fm}.
Furthermore, the broad ridge separated from the main peak denotes the non-Gaussianity
of $G_s(r,\theta; t)$.
A tagged molecule is trapped by a cage composed of neighbor molecules
for longer times in supercooled regime.
The rotational relaxation time $\tau_1$, of which the characteristic angle
is $\pi/2$ rad, is governed by this
large rotational mobility.
The single molecule eventually begins diffusion by escaping from the cage, utilizing
large translational and rotational mobilities.
Thus, $\tau_{1}$ is regarded as the time scale coupled with 
$D_\mathrm{t}$.
In addition, the time scale of $\tau_1$ is similar to the hydrogen-bond
lifetime $\tau_\mathrm{HB}$, as demonstrated in Fig.~\ref{fig2}(a).

In contrast, the relaxation time $\tau_6$ corresponds to a molecular reorientation
with an angle of 0.37 rad, which lies near to the dominant peak of
$G_s(r, \theta; t)$ at $t=10$ ns (see Fig.~\ref{fig5}(c)).
The higher-order $\ell$ mostly highlights immobile molecules both for
translational and rotational motions, which will contribute to the dynamical
heterogeneities.
%In other words, the relaxation of higher-order $\ell$ is governed by dynamic
%heterogeneities.
To investigate this, we characterize
the dynamic heterogeneity of translational and rotational motions 
using the four-point correlation functions, $\chi_\mathrm{t, r}(t)$ (see
Methods and Supplementary Fig.~S3(a)).
We found that the time scale $\tau_6$ is akin to the peak time of $\chi_\mathrm{t,
r}(t)$, which shows that its temperature dependence is similar to that of 
$\tau_\alpha$ (see Supplementary Fig.~S3(b)).
Consequently, the similar temperature dependences between
$\tau_6$ and $\tau_\alpha$ are demonstrated in Fig.~\ref{fig2}(a).

Finally, we discuss the strong decoupling behavior of the translational and
rotational diffusion constants, as demonstrated in Fig.~\ref{fig1}.
As already pointed out in Ref.~\citenum{Laage:2008he}, 
the use of the rotational diffusion constant $D_\mathrm{r}$ needs
particular care due to the limitation of the angular Brownian motion scnenario.
Furthermore, it has been revealed that $D_\mathrm{r}$ is superfacial
for describing the reorientational motion in supercooled molecular
liquids.~\cite{Chong:2009ci}
The angular mean-square displacement $\ave{\Delta \phi(t)^2}$ 
is largely influenced by the accumulation of the libration motion,
which has a time scale of 0.1 ps.
Each molecule can rotate, despite being trapped by the cage,
at this short time scale, as indicated in Fig.~\ref{fig5}(a).
Accordingly, the angular mean-square displacement 
exhibits a plateau, but its persistent time is much
smaller than that of $C_\ell(t)$, particularly at lower temperatures (See 
Supplementary Figs.~S1 and S2).
In contrast, the plateau of $C_\ell(t)$ after the time scale of libration
motion indicates the occurrence of the cage effect (see Supplementary Fig.~S2).
These findings imply that the decoupling between $D_\mathrm{t}$ and
$D_\mathrm{r}$ has no direct relevance to the real
translational-rotational coupling, $D_t\propto \tau_1^{-1}$.
This translational-rotational coupling scenario is in accord with the
observasion in supercooled molecular liquids.~\cite{Chong:2009ci}

\section*{Discussion}
In this paper, we report the numerical results of MD simulations of the
relationship between the translational and rotational dynamics in TIP4P/2005
supercooled liquid water by performing MD simulations.
Our contributions to the assessment of translational and rotational SE relations and
the Debye model can be summarized as follows:

(i) Both translational and rotational SE relations, $D_\mathrm{t}\propto
(\eta/T)^{-1}$
and $D_\mathrm{r}\propto (\eta/T)^{-1}$, are significantly violated in supercooled states. 
In particular, the rotational SE relation is 
violated stronger to a greater extent than that of translational SE relation.
Correspondingly, the rotational hydrodynamic radius becomes significantly smaller than 
translational one with decreasing temperature.

(ii) We test the validity of the Debye model, $D_\mathrm{r}\propto {\tau_\ell}^{-1}$,
for the orders $\ell=1$, $2$, $3$, and $6$ of the Legendre polynomials, demonstrating
that the rotational relaxation time $\tau_\ell$
is entirely inconsistent with the rotational diffusion constant $D_\mathrm{r}$.

(iii) Furthermore, we systematically examine the SED
relations $\tau_\ell T/\eta$ and $D_\mathrm{t}\tau_\ell$.
We reveal
that these SED relations strongly depend
on the order of $\ell$, leading to the following spurious argument: The SED
relation $\tau_\ell\propto \eta/T$ is violated
with the lowest-order rotational relaxation time $\tau_1$, but is instead 
at the higher-order time scale of $\tau_6$.
Instead, we find that
$D_\mathrm{t}\tau_6$ deviates from values at high temperatures,
similarly to the violation of the translational SE relation $D_\mathrm{t}\eta/T$, while
$D_\mathrm{t}\tau_1$ superficially satisfies the SED relation.

(iv) We observe the coupling between 
the translational diffusion constant, $D_\mathrm{t}$, and the lowest
rotational relaxation time, $\tau_1$.
We characterize 
the correlation between large translational and rotational mobilities using
from the van Hover correlation function $G_s(r, \theta; t)$.
Furthermore, we find that $\tau_1$ exhibits the temperature dependence similar to
that of the hydrogen bond breakage time $\tau_\mathrm{HB}$, which is
consistent with the previously demonstrated result,
$D_\mathrm{t}\propto{\tau_\mathrm{HB}}^{-1}$.~\cite{Kawasaki:2017gw}

(v)
On the contrary, we show that the higher-order rotational relaxation
time $\tau_6$ is analogous with the
$\alpha$-relaxation time $\tau_\alpha$, rather than with $\tau_\mathrm{HB}$.
This time scale is characterized by immobile molecular mobilities
showing dynamic heterogeneities, which we investigate using
the four-point correlation functions $\chi_\mathrm{t}(t)$ and
$\chi_\mathrm{r}(t)$.
It is also of essential to examine the role of the length scale of
dynamic heterogeneities $\xi_4$ on the violations of the SE/SED
relations in supercooled water.
The $\xi$ value is conventionally quantified by the wave-number
dependence of the four-point correlation
functions.~\cite{Toninelli:2005ci}
This calculation necessitates MD simulations using more substantial
large systems, which are currently undertaken.

(vi)
In conclusion, in this paper we provide significant and unprecedented
insights into the appropriate assessment of SE, Debye, and SED
relationships, in doing so clarifying previously awkward and confusing contradictions.
Finally, it is worth mentioning the importance of the density
dependence on the SE/SED relations in supercooled water.
In fact, both $\eta$ and $D_\mathrm{t}$ show
anomalous density dependence, particularly at low temperaures.~\cite{MonterodeHijes:2018ec}
Further investigations along this line are necessary to clarify this issue.

\section*{Methods}

\subsection*{Molecular dynamics simulations}
We performed MD simulations of liquid water using the Large-scale
Atomic/Molecular Massively Parallel Simulator
(LAMMPS)~\cite{Plimpton:1995wl}, and used
the TIP4P/2005 model to calculate the water molecule potentials.~\cite{Abascal:2005ka}
Other MD simulations were carried out to investigate various properties in supercooled
states of this model.~\cite{Abascal:2010dw, Sumi:2013fy, Overduin:2013cu,
DeMarzio:2016hl, Hamm:2016hf, Singh:2016bu, Gonzalez:2016gr,
Handle:2018cn, MonterodeHijes:2018ec, Saito:2018dn, Saito:2019ht}
The comparison with other rigid non-poralizable models was also made
in the recent review.~\cite{Vega:2011bb}
Remark that recent ab initio MD simulations provide a more realistic behavior of the
dynamics in supercooled regime.~\cite{Morawietz:2016cp}
We used a Coulombic cutoff 1 nm.
The particle-particle particle-mesh solver was utilized to calculate
long-range Coulomb interactions, and
the SHAKE algorithm was also used for bond and angle constraints.
Periodic boundary conditions were used, and the time step of simulation was 1 fs.
First, we employed the $NVT$ ensembles for $N = 1,000$ water molecules was employed at various temperatures ($T =
300, 280, 260, 250, 240, 230, 220, 210, 200$, and 190 K) with a fixed mass
density of $\rho = 1$ g cm$^{-3}$. 
The corresponding system size was $L = 31.04$ \AA. 
We conducted the $NVE$ ensemble simulations 
after the equilibration with a sufficient long time at each temperature,
The dynamical quantities including, the $\alpha$-relaxation time $\tau_\alpha$,
the translational diffusion constant $D_\mathrm{t}$, and the hydrogen-bond
lifetime $\tau_\mathrm{HB}$, and shear viscosity $\eta$ used in here are
reported in a previous study.~\cite{Kawasaki:2017gw}
In this study, we newly calculated time correlation functions
for characterzing the rotational diffusion constant $D_\mathrm{r}$, the
rotational relaxation time of the $\ell$-th degree Legendre polynomials $\tau_\ell$.
Furthermore, the four-point correlation function was also calculated for
characterizing dynamic heterogeneities of rotational molecular motions.
The trajectories for the calculations of various quantities were for 10
ns ($T \ge 220$ K) and 100 ns ($T \le 210$ K).
We average over five independent simulation runs for the calculations.

\subsection*{Rotational diffusion constant and rotational relaxation
  time}

We calculated the angular mean-square displacement
$
\ave{\Delta \phi (t)^2} = (1/N)\sum_{j=1}^N \ave{|\Delta \vec{\phi}_j (t)|^2},
$
following Ref.~\citenum{Mazza:2007kr} (see Supplementary Fig.~S1(a)).
We obtained 
the angular displacement vector $\Delta \vec{\phi}_j (t)$ of
the molecule $j$ is obtained through the time
integration of the angular velocity vector, 
$
\vec{\phi}_j(t) = \int_{t'}^{t'+t}  \vec{\omega}_j (t) \mathrm{d}t,
$
where the angular velocity vector $\vec{\omega}_j (t)$ of the molecule $j$
is given by the cross-product of the normalized polarization vector
$\vec{e}_j(t)$ as,
$
\vec{\omega} _j(t)= \vec{e}_j(t)\times \vec{e}_j(t + \Delta t)/\Delta t, 
$
with the magnitude $|\vec{\omega}_j (t)|=\cos^{-1}(\vec{e}_j(t)\cdot \vec{e}_j(t + \Delta t))$.
Note that $\Delta t$ is chosen by a sufficiently small time interval; 
this was 10 fs in our calculations.
We determined the 
rotational diffusion constant $D_\mathrm{r}$ was determined from the long-time
limit of $\ave{\Delta \vec{\phi} (t)^2}$ as
$
D_\mathrm{r}=\lim_{t\to \infty}\ave{\Delta \phi (t)^2}/4t. 
\label{eq:Mazza}
$
Furthermore, we independently calculated the angular velocity time correlation
function,
$
C_{\Omega}(t) = (1/3N)\sum_{j=1}^N\ave{\vec{\Omega}_j(t) \cdot \vec{\Omega}_j(0)},
$
where $\vec{\Omega}_j(t)$ denotes the angular velocity vector of the molecule $j$ in
the world reference frame following Ref.~\citenum{Rozmanov:2012ja} (see Supplementary Fig.~S1(b)).
We also used the 
Green--Kubo formula to obtain $D_\mathrm{r}$ as
$
D_\mathrm{r} =\int_0^{\infty} C_{\Omega} (t) \mathrm{d}t.
$
We confirmed that the $D_\mathrm{r}$ values obtained from these two methods
are consistent at any temperature.

The rotational correlation functions
$C_{\ell}(t)$ is defined by the
autocorrelation function of the normalized polarization vector $\vec{e}_j(t)$
as 
$
C_{\ell}(t) = (1/N)\sum_{j=1}^N \ave{P_{\ell}[\vec{e}_j(t) \cdot \vec{e}_j(0)]}, 
$
where $P_{\ell}[x]$ is the $\ell$-th order Legendre polynomial as a function of $x$
(see Supplementary Fig.~S2).
$C_{\ell}(t)$ decays from 1 to 0 as $t$ 
increases. 
We obtained the $\ell$-th ($\ell=1$, $2$, $3$, and $6$) order rotational
relaxation time $\tau_{\ell}$
by fitting $C_{\ell}(t)$ to the Kohlrausch--Williams--Watts function
$A\exp{\{-(t/\tau_{\ell})^{\beta_{\ell}}\}}$.

\subsection*{Rotational four-point correlation functions}

We used the four-point correlation function to elucidate
the degree of dynamic heterogeneity in supercooled
liquids.~\cite{Toninelli:2005ci}
The four-point correlation $\chi_\mathrm{t}(t)$ for translational
motions is defined by the variance of the intermediate scattering
function $F_s(k, t)$ as,
$
\chi_\mathrm{t}(t) = N\left[\langle \hat F_s(k, t)^2\rangle -\langle
		\hat F_s(k, t)\rangle^2 \right], 
$
with $\hat F_s(t)=(1/N)\sum_{j=1}^N\cos[\vec{k}\cdot \Delta
\vec{r}_j(t)]$.
We previously calculated $\chi_\mathrm{t}(t)$ using the wave-number $k=3.0$ \AA$^{-1}$,
and quantified the peak time $\tau_\mathrm{t}$ (note that
the same quantity was denoted by $\tau_\mathrm{\chi_4}$ in Ref.~\citenum{Kawasaki:2017gw}).
The rotational four-point correlation function
$\chi_\mathrm{r}(t)$ can be analogously defined as
$
\chi_\mathrm{r}(t) = N\left[\langle \hat C_\ell(t)^2\rangle -\langle \hat C_\ell(t)\rangle^2\right],
$
with $\hat C_\ell(t) = (1/N)\sum_{j=1}^N P_\ell[\vec{e}_j(t)\cdot\vec{e}_j(0)]$.
The peak time of $\chi_\mathrm{r}(t)$ is represented by $\tau_\mathrm{r}$.

\acknowledgments
We thank K. Miyazaki, N. Matubayasi, T. Nakamura, and H. Shiba for 
valuable discussions. 
This work was partly supported by JSPS KAKENHI Grant Numbers JP15H06263,
JP16H06018, JP18H01188, and JP19K03767.
The numerical calculations were performed at the Research Center of
Computational Science, Okazaki, Japan.

\clearpage
\widetext

\setcounter{equation}{0}
\setcounter{figure}{0}
\setcounter{table}{0}
\setcounter{page}{1}

\noindent{\bf\Large Supplementary Material}
%\vspace{2mm}
%\begin{center}
%{\bf  \large  The relationship between the rotational and translational relaxations in supercooled water }
%\\
% 
%{Takeshi  Kawasaki$^1$ and Kang Kim$^2$}\\
%$^1$Department of Physics, Nagoya University, Nagoya 464-8602, Japan\\
%$^2$Division of Chemical Engineering, Graduate School of Engineering Science, Osaka University, Osaka 560-8531, Japan
%\end{center}

\setcounter{equation}{0}
\renewcommand{\theequation}{S.\arabic{equation}}
\renewcommand{\thefigure}{S\arabic{figure}}

\renewcommand{\bibnumfmt}[1]{[S#1]}
\renewcommand{\citenumfont}[1]{S#1}

%S1
\begin{figure*}[h]
\includegraphics[width=0.65\textwidth]{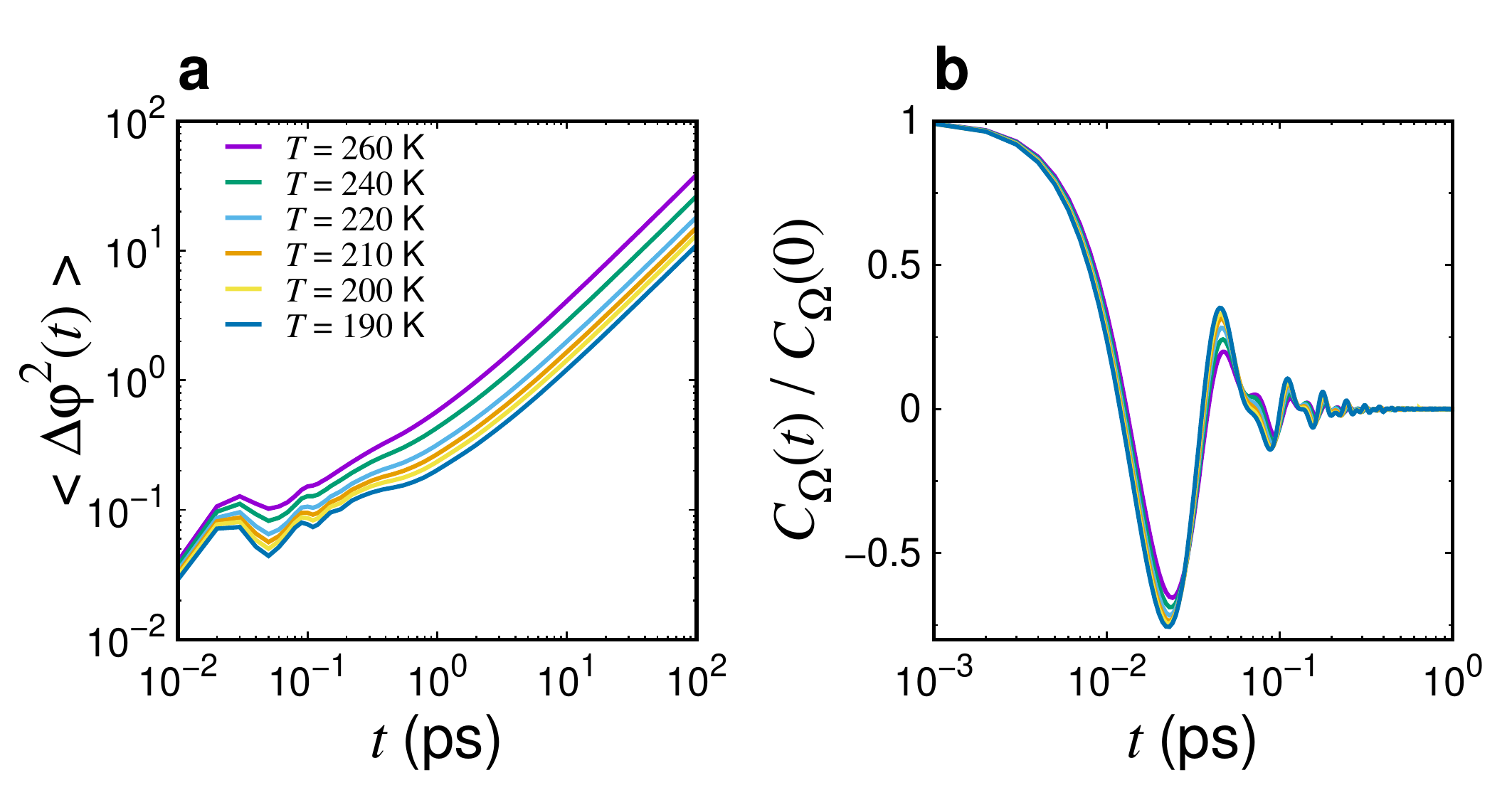}
\caption{\label{figS1} 
(a) Mean squared angular displacement $\ave{\Delta \phi(t)^2}$.
(b) Angular velocity correlation function $C_{\Omega} (t)$.
}
\end{figure*}

%S2
\begin{figure*}[h]
\includegraphics[width=0.65\textwidth]{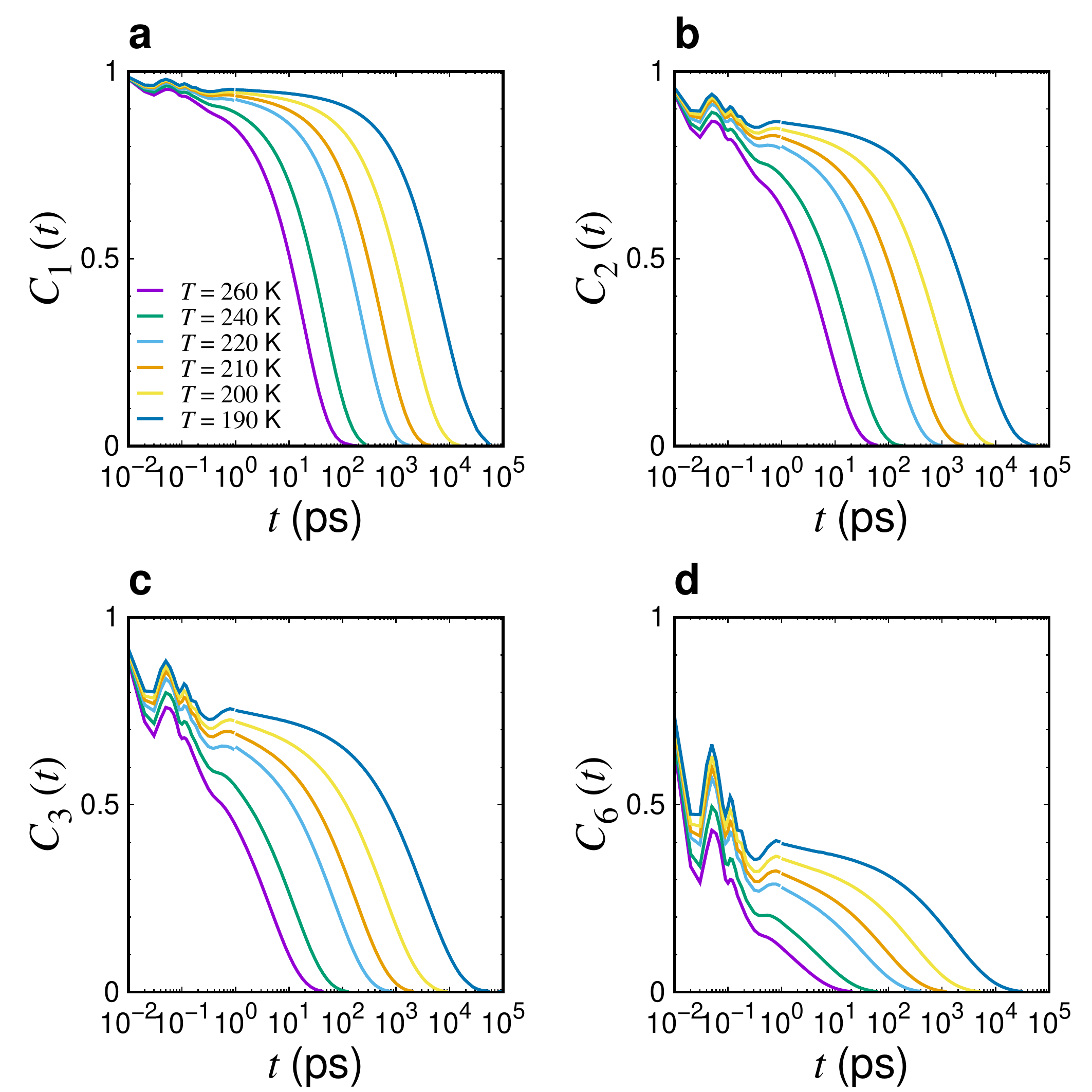}
\caption{\label{figS2} 
Rotational correlation function $C_{\ell}(t)$ for $\ell=1$ (a), $2$
 (b), $3$ (c), and $6$ (d). 
}
\end{figure*}

%S3
\begin{figure*}[h]
\includegraphics[width=0.65\textwidth]{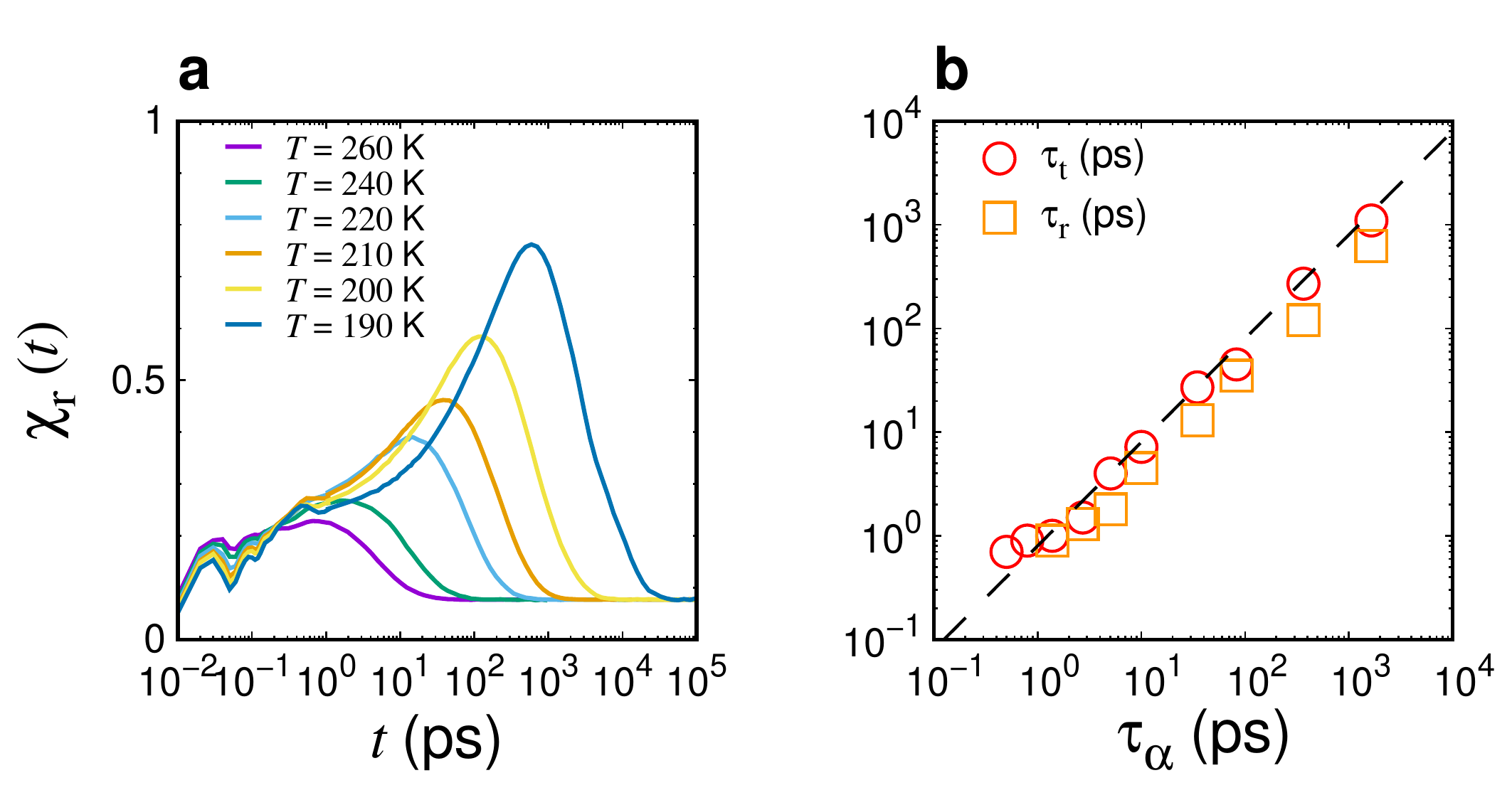}
\caption{\label{figS3} 
(a) Rotational four-point correlation functions,
 $\chi_\mathrm{r}(t)$ with the order $\ell=6$.
(b) The relationship between $\tau_\mathrm{t}$ and $\tau_\mathrm{r}$,
 and $\alpha$-relaxation time $\tau_\alpha$, as
 determined by the intermediate scattering function $F_s(k, t)$, with
 $k=3.0$~\AA$^{-1}$.
}
\end{figure*}

\end{document}